\documentclass[letterpaper]{article} 
\usepackage{aaai25}  
\usepackage{times}  
\usepackage{helvet}  
\usepackage{courier}  
\usepackage[hyphens]{url}  
\usepackage{graphicx} 
\usepackage{natbib}  
\usepackage{caption} 
\usepackage{algorithm}
\usepackage{algorithmic}

\usepackage{newfloat}
\usepackage{listings}
\usepackage{xcolor}
\DeclareCaptionStyle{ruled}{labelfont=normalfont,labelsep=colon,strut=off} 
\floatstyle{ruled}
\newfloat{listing}{tb}{lst}{}
\floatname{listing}{Listing}
\title{Textured Mesh Saliency: Bridging Geometry and Texture \\ for Human Perception in 3D Graphics}
\author {
    Kaiwei~Zhang\textsuperscript{\rm 1},
    Dandan~Zhu\textsuperscript{\rm 2}$^*$,
    Xiongkuo~Min\textsuperscript{\rm 1}$^*$,
    Guangtao~Zhai\textsuperscript{\rm 1}\thanks{Corresponding author.}
}
\affiliations {
    \textsuperscript{\rm 1}Shanghai Jiao Tong University\\
    \textsuperscript{\rm 2}East China Normal University\\
    zhangkaiwei@sjtu.edu.cn, ddzhu@mail.ecnu.edu.cn\\
    \textcolor[rgb]{1.0,0.1,0.5}{https://github.com/kaviezhang/TexturedMeshSaliency}
}

\usepackage{bibentry}

\begin{document}

\maketitle

\begin{abstract}
Textured meshes significantly enhance the realism and detail of objects by mapping intricate texture details onto the geometric structure of 3D models. This advancement is valuable across various applications, including entertainment, education, and industry. While traditional mesh saliency studies focus on non-textured meshes, our work explores the complexities introduced by detailed texture patterns. We present a new dataset for textured mesh saliency, created through an innovative eye-tracking experiment in a six degrees of freedom (6-DOF) VR environment. This dataset addresses the limitations of previous studies by providing comprehensive eye-tracking data from multiple viewpoints, thereby advancing our understanding of human visual behavior and supporting more accurate and effective 3D content creation. Our proposed model predicts saliency maps for textured mesh surfaces by treating each triangular face as an individual unit and assigning a saliency density value to reflect the importance of each local surface region. The model incorporates a texture alignment module and a geometric extraction module, combined with an aggregation module to integrate texture and geometry for precise saliency prediction. We believe this approach will enhance the visual fidelity of geometric processing algorithms while ensuring efficient use of computational resources, which is crucial for real-time rendering and high-detail applications such as VR and gaming.
\end{abstract}

%

\section{Introduction}
\begin{figure}[t]
  \centering
  \includegraphics[width=0.4\textwidth]{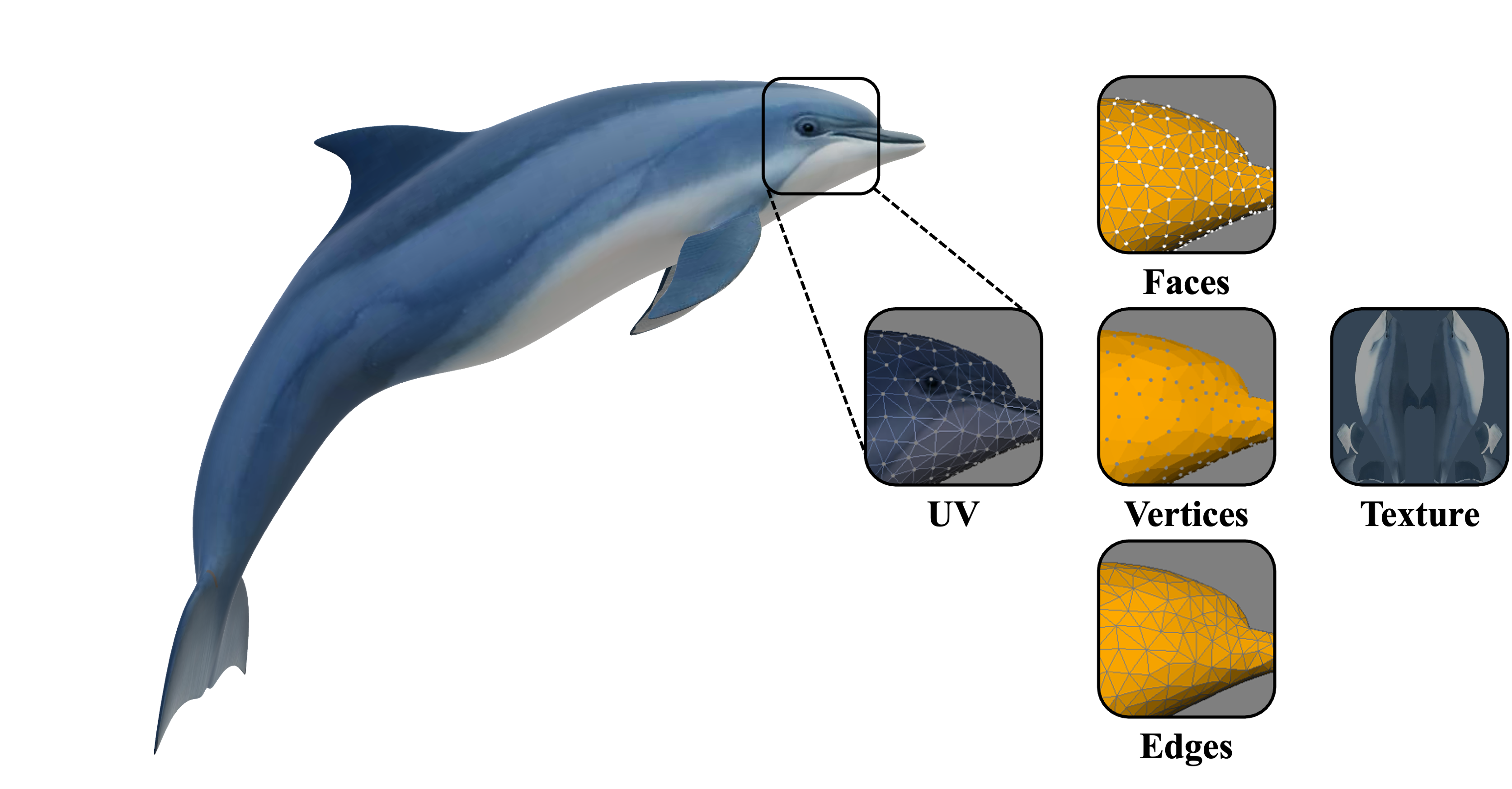}
  \setlength{\abovecaptionskip}{-0pt}
  \caption{In textured meshes, a texture is a 2D image that, through UV mapping, is accurately projected onto the faces of the mesh, adding visual details and realism. UV coordinates define each vertex's position on the texture image, ensuring that the texture is correctly mapped onto the 3D surface of the mesh.}
\label{fig:intro}
\end{figure}

Situated at the intersection of computer graphics and computer vision, textured meshes bring a higher level of visual coherence and realism to 3D content.
Mesh topology defines the underlying geometry and structure, while the texture contributes intricate details such as colors, patterns, and material properties.
As illustrated in Figure \ref{fig:intro}, the method of mapping texture pixels onto the geometric surface polygons enhances the shading details, giving the model a more vibrant appearance and making objects look more lifelike.
By combining geometric data with texture mapping, textured meshes deliver detailed, realistic, and efficient representations of 3D objects, making them valuable in applications such as entertainment, education, and industry.

In human-centered visual applications, techniques like mesh simplification \cite{cignoni1998comparison} and smoothing \cite{jones2003non} pose significant challenges to visual perception \cite{zhang2023synergetic}.
These challenges are especially pronounced in applications that demand real-time rendering and high levels of detail, such as VR and gaming environments.
Particularly in game engines, Level of Detail (LOD) \cite{luebke2003level} technology adjusts the complexity of mesh models based on varying distances, crucial for maintaining visual quality while minimizing computational burden in large-scale scenes with multiple models on screen.
By integrating human perceptual insights into mesh processing, visual saliency provides a reference for the importance of different regions within a mesh. Consequently, regions identified as more significant can be optimized with finer details, while fewer resources are allocated to less critical regions. This approach not only enhances visual fidelity but also ensures efficient use of computational resources, leading to smoother performance and better user experiences.
Thus, a thorough investigation into the saliency of textured meshes not only enhances our understanding of users' visual behavior patterns but also drives optimization in virtual environment design, enhances user experience, and supports data-driven content creation.


Existing mesh saliency techniques primarily focus on non-textured meshes \cite{lee2005mesh,martin2024sal3d}. Recently, colored meshes have also been treated as the subject of mesh saliency in \cite{ding2023towards}, with the assumption that vertex color is a significant factor influencing visual attention. However, the coloring method for colored meshes merely stores color information at the mesh vertices, which lacks the capacity to represent complex patterns and textures. This vertex coloring technique is more commonly used in point cloud data and is suitable only for basic visualization tools, limiting its practical applications.
In contrast, textured meshes dominate high-quality rendering applications such as games and animation. The use of intricate texture images to create sophisticated and realistic visual effects presents entirely different challenges for mesh saliency in textured meshes.
Clearly, extending the understanding and utilization of saliency to textured meshes can have a more significant impact on enhancing the rendering quality and visual performance of 3D scenes.

Therefore, we develop a novel experimental procedure and create the first saliency dataset for textured meshes. This experiment ensures that participants could comprehensively collect eye-tracking data while observing 100 textured mesh models from any viewpoint within a 6-DOF VR environment.
It addresses the issues of incomplete and less reliable eye-tracking data caused by uneven lighting conditions and limited viewing directions in virtual environments.
Furthermore, based on experimental analysis, the color patterns and geometric structures of textured meshes should be considered as synergistic factors attracting the human visual system. Existing deep learning methods for geometric processing focus on non-textured meshes, which can lead to unpredictable confusion in visual attention inference for textured models. To address this, we introduce a saliency prediction model specifically designed for textured meshes. This model uses implicit interpolation to align irregularly shaped texture features with geometric features, providing a more accurate representation of surface importance in textured meshes.
Meanwhile, it establishes a new paradigm for deep learning-based geometric processing of textured meshes.
In summary, the key contributions of our work are as follows:

\begin{itemize}
\item We establish a new textured mesh saliency dataset based on a novel eye-tracking experimental design. This dataset fills the gap in human perceptual insights for high-quality rendering applications, enhances our understanding of user visual behavior, and can also drive data-driven 3D AI-generated content creation.
\item To fully explore the synergistic effect of texture and geometry on saliency distribution, we propose a texture-implicitly aware saliency prediction model for textured meshes. This model employs implicit interpolation through UV mapping to align irregular 2D texture patterns with 3D structures.
\item We design experiments to evaluate model performance with only geometric features and then incorporate vertex colors and texture mapping to assess the impact of each feature. It reveals the contribution of each feature to saliency prediction and demonstrates how texture improve prediction accuracy.
\end{itemize}

\begin{figure*}[ht]
  \centering
  \includegraphics[width=0.90\textwidth]{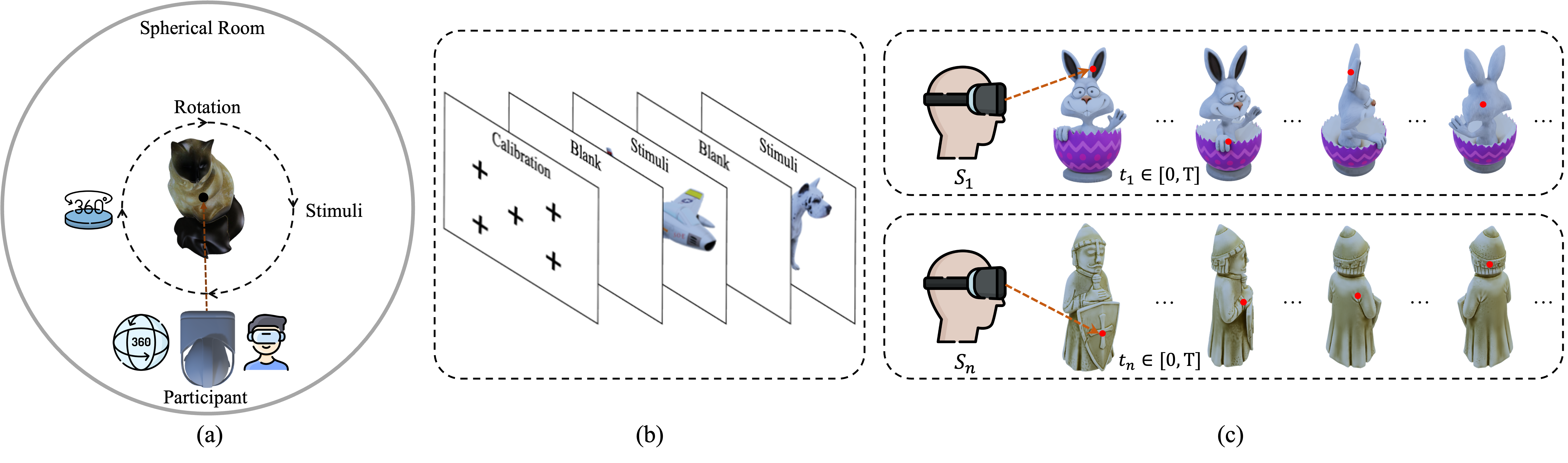}
  \setlength{\abovecaptionskip}{-0pt}
  \caption{VR eye-tracking experiment for textured meshes. (a) illustrates the experimental setup, where the textured meshes are placed at the center of a spherical space and rotated clockwise. (b) shows the sequence of content displayed during the data collection process. (c) presents the recording process of fixation points during the mesh rotation.}
\label{fig:proce}
\end{figure*}

\section{Related Work}
\subsection{Saliency Datasets}
Recently, there has been an increase in proposed mesh saliency datasets. Initially, the studies like \cite{kim2010mesh,lavoue2018visual} utilized projected 2D views of 3D mesh models as visual stimuli and capture gaze points with on-screen eye trackers. These 2D coordinates are then mapped onto the 3D model. A Gaussian filter is subsequently applied to smooth the fixation points, assigning a fixation density value to each vertex.
In some following studies \cite{ding2023towards, martin2024sal3d}, eye-tracking data for meshes were collected within a VR environment. The mesh model is positioned at the center of a virtual room. During the experiment, each participant is initially seated, facing a different viewing orientation. The viewing direction is then rotated clockwise by 30° every 10 seconds to comprehensively cover a wide range of viewing angles. This experimental setup yields a small-scale mesh saliency dataset.
Although these 3D mesh saliency datasets are valuable, their data collection methods have various shortcomings, and the scale of eye-tracking datasets in VR environments is also limited. We improve the VR experimental design by enhancing lighting conditions and refining the eye-tracking data collection process, resulting in the creation of a new textured mesh saliency dataset.

\subsection{Saliency Prediction}
With the advancement of 3D mesh models, there is an urgent need for research in mesh saliency prediction. The concept of mesh saliency was first introduced by Lee et al. \cite{lee2005mesh}, who employed a center-surround operation on the Gaussian-weighted mean curvature to compute mesh saliency.
In contrast to the aforementioned methods, the research presented by Song et al. \cite{song2021mesh} employed a weakly supervised approach. This method involves training on image saliency and object category labels to predict saliency maps for projected 2D views from various viewpoints. Subsequently, a 2D-3D saliency mapping method is used to generate heatmaps for mesh saliency.
Similarly, Abid et al. \cite{abid2020towards} used a 2D saliency detection model trained on the SALICON dataset to generate projected saliency results. 
In \cite{martin2024sal3d}, a deep learning-based saliency prediction model was build upon a state-of-the-art classification network.
Despite these advancements, no methods currently predict saliency for textured meshes. To address this gap, we propose a new prediction network that integrates texture mapping with geometric structure.


\section{Dataset Creation}
\subsection{Data Collection}
VR environments provide innovative solutions to challenges that are difficult to replicate in real-life settings. Previous research \cite{sitzmann2018saliency} has explored the use of VR headsets and eye-tracking devices to capture head movements and gaze data of participants observing omnidirectional videos, enabling studies on visual saliency in immersive contexts. Therefore, we propose a novel VR spatial experiment design aimed at collecting eye-tracking data for textured mesh saliency. Due to space constraints, a detailed description of the dataset is provided in Appendix.

\subsubsection{Stimuli.}
To better analyze the intrinsic relationships in saliency distribution, it is crucial to collect meshes that feature rich texture colors and consistent geometric styles as visual stimuli. We select a total of 100 textured mesh models from the Free3d \cite{Free3d} and TurboSquid \cite{TurboSquid} repository for eye-tracking experiments.

\subsubsection{Participants.}
A total of 30 participants (14 males and 16 females aging between 18 and 35) actively participate in our eye-tracking experiment. All participants are all newcomers to the field of computer graphics and report normal or corrected-to-normal vision, along with normal color perception.
Before viewing the stimuli, participants undergo training and eye-tracking calibration. Only participants who successfully complete this step are included in the experiment.

\subsubsection{Procedure.}
Participants in the eye-tracking experiment wear the HTC Vive Pro Eye device and observe the displayed content while they could move freely within a $3\times3$ meter space. As shown in Figure \ref{fig:proce}(a), a spherical space with a 5-meter radius is built to ensure consistent and smooth backgrounds. We use uniform area lighting to eliminate uneven illumination interference.
All textured meshes are presented sequentially in a random order and rotates clockwise at a speed of 15 degrees per second to make subjects observe the overall structure and detailed areas from comprehensive angles.
Figure \ref{fig:proce}(b) illustrates the entire process flow of eye-tracking data collection.
Throughout the observation process, an eye-tracking script records eye movement data at each moment, including gaze origin, gaze direction, head position, and head direction.

\begin{figure}[!b]
  \centering
  \includegraphics[width=0.4\textwidth]{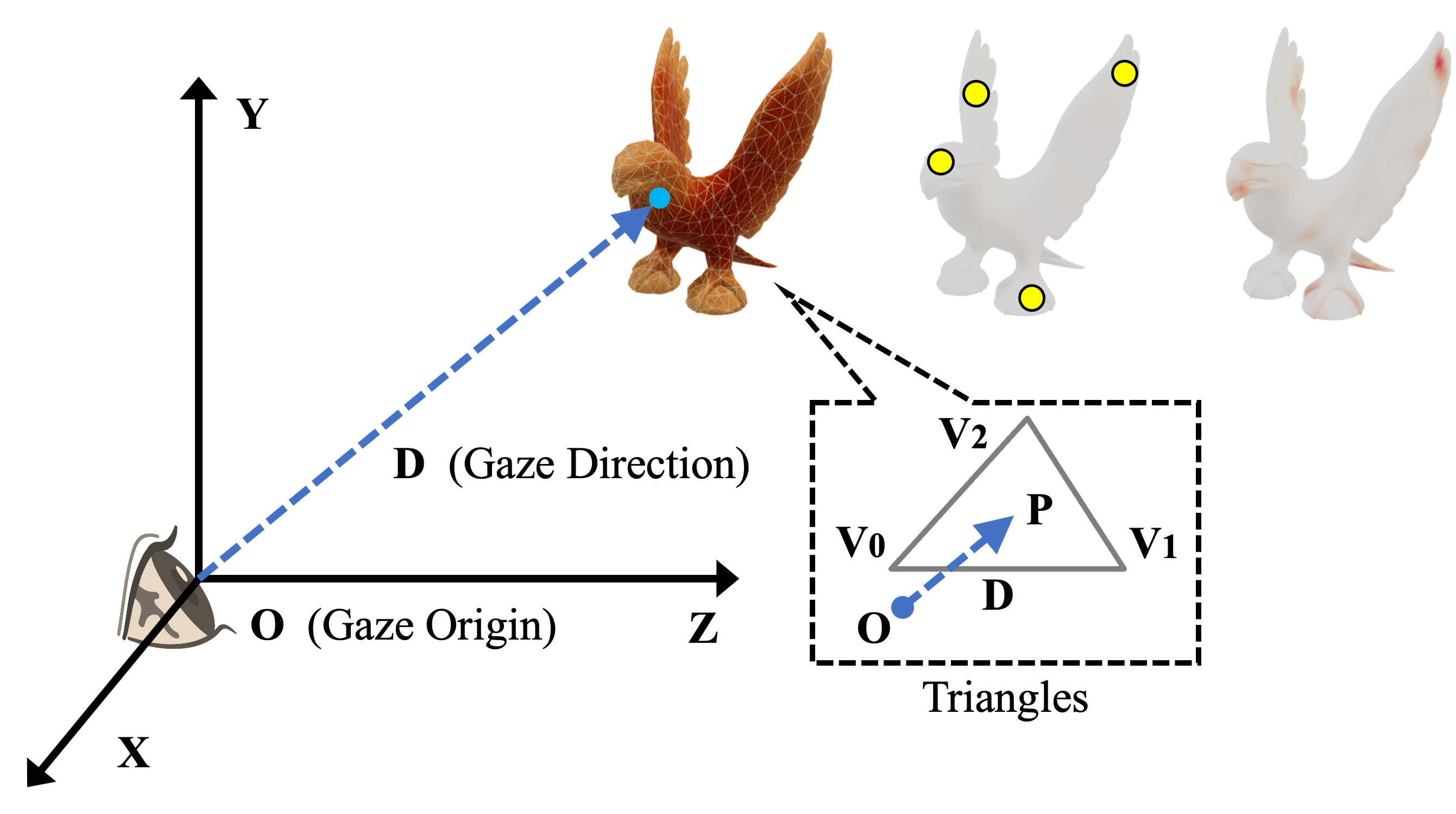}
  \setlength{\abovecaptionskip}{-0pt}
  \caption{The intersection between the gaze ray and the textured mesh model. The yellow dots represent the gaze fixation triangles where the gaze lingers on the model. The saliency map is presented in the form of a heatmap.}
\label{fig:inter}
\end{figure}

\subsection{Data Processing}
To detect the intersections between eye-tracking gaze rays and the texture mesh models, we utilize the Barycentric Interpolation algorithm and the Möller–Trumbore algorithm \cite{moller2005fast} to mathematically model the intersection process between the triangles on the mesh surface and the gaze rays, as illustrated in Figure \ref{fig:inter}. By applying Cramer's rule and acceleration structures like Bounding Volume Hierarchies (BVH) \cite{gunther2007realtime}, we can efficiently narrow down the set of triangles to be checked, thus speeding up the computation of all gaze intersections. The gaze intersections are then classified using the velocity-based I-VT algorithm \cite{salvucci2000identifying}, which filters out saccades and other noise.
Finally, a cone-based Gaussian kernel is applied to smooth the remaining fixations, generating a saliency map for the textured mesh.

\subsection{Data Analysis}
Similar to the methods used in visual preference research for mesh saliency, we select surface Gaussian curvature and texture patch color variance as quantitative measures of local geometric features and color differences.
Using statistical methods, we conduct random sampling analysis on salient and non-salient regions based on these two metrics. We repeat the statistical process 100 times and randomly sample 1000 points on each model. The results show that 87.18\% of the salient triangles have higher Gaussian curvature values compared to non-salient triangles. Additionally, 82.17\% of the salient triangles exhibit greater color texture variance than non-salient triangles. Therefore, on textured meshes, surface curvature and texture richness attract more visual preference.

\begin{figure*}[!htbp]
  \centering
  \includegraphics[width=0.95\textwidth]{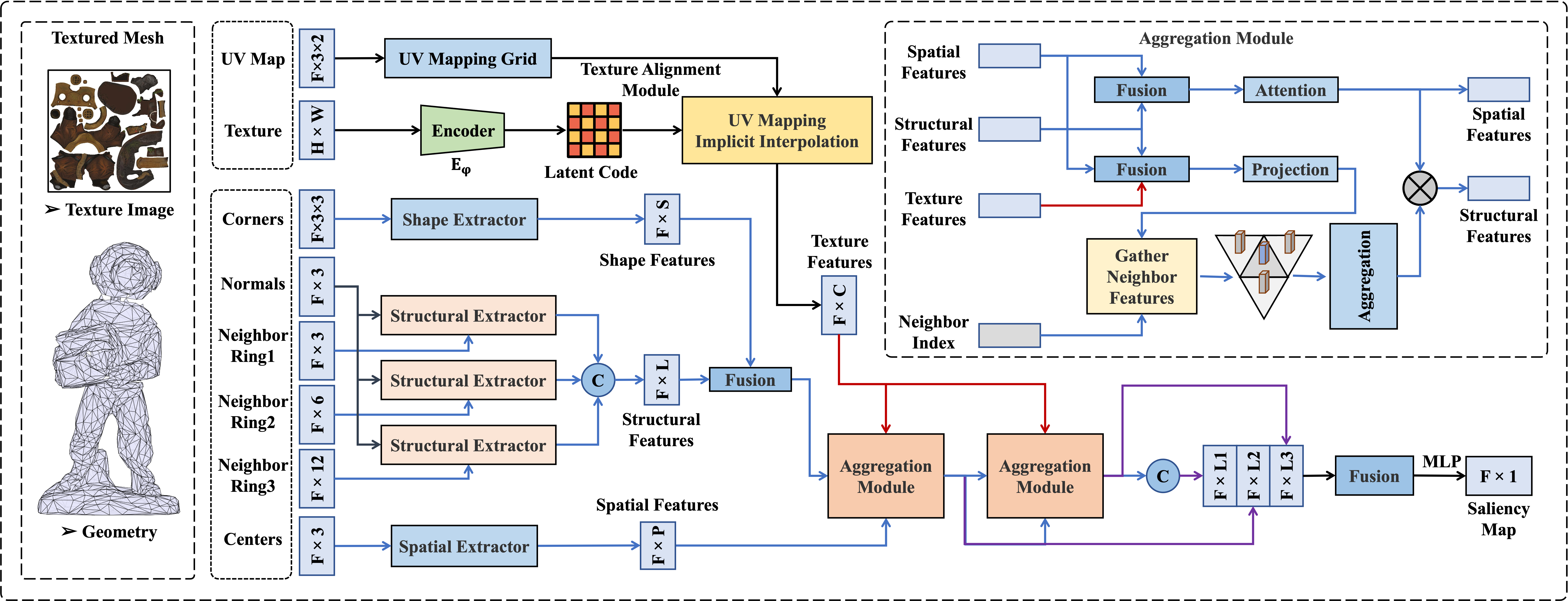}
  \setlength{\abovecaptionskip}{4pt}
  \caption{Model architecture. The network for texture and UV map inputs is the texture alignment module, while the network that processes the geometry inputs is the geometric extraction module.}
\label{fig:mod}
\end{figure*}

\begin{figure}[!b]
  \centering
  \includegraphics[width=0.45\textwidth]{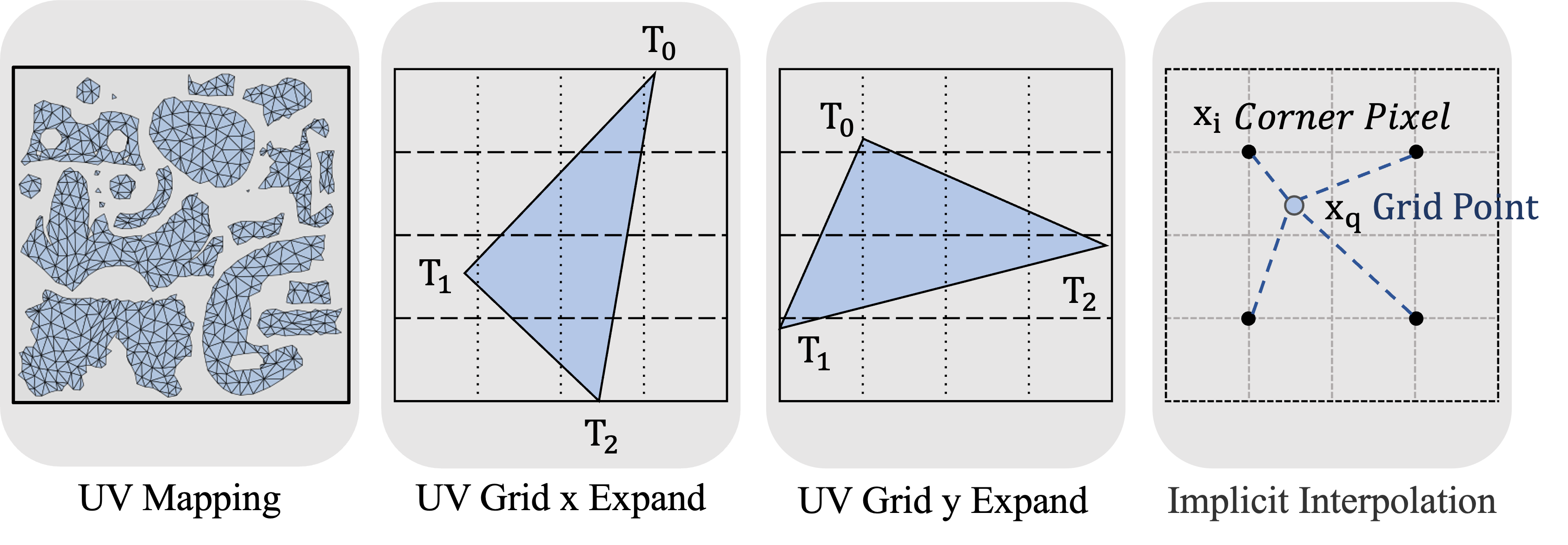}
  \setlength{\abovecaptionskip}{-6pt}
  \caption{UV mapping grid and implicit interpolation for texture feature alignment.}
\label{fig:grid}
\end{figure}

\section{Proposed Model}


In this section, we design a model for predicting saliency maps on the surface of textured meshes. The saliency prediction task for the mesh operates at the level of triangular faces, providing a saliency density value for each face to reflect the importance of local surface regions.
As illustrated in Figure \ref{fig:mod}, the model primarily consists of the texture alignment module, the geometric extraction module, and the aggregation module that integrates texture features with geometric structure to ultimately predict the saliency map.
To better introduce our model, the missing details are provided in Appendix.

\subsection{Texture Alignment Module}
The texture alignment module extracts features from the texture image and maps the feature vectors to the triangular faces according to their corresponding UV coordinates.
The texture image is typically input as an image tensor with dimensions $H\times W$, where $H$ and $W$ represent the height and width of the image, respectively, and the RGB channel count is 3. UV texture coordinates are used to correctly map the 2D texture image onto the vertices of the 3D model. These are input as an $F\times3\times2$ matrix, where $F$ is the number of triangular faces, and each vertex of a face has two texture coordinates $(u, v)$. In the original format, shared vertices between adjacent triangular faces may have different corresponding UV values due to their different face contexts.

\subsubsection{Latent Code}
Inspired by implicit neural representations \cite{chen2021learning, zhang2022implicit}, the latent code map provides a shared continuous implicit representation for both the texture image domain and the feature domain. This ensures that the pixel positions in the texture image corresponding to the UV coordinates align with the positions in the high-dimensional feature map.
In Figure \ref{fig:mod}, the texture image is represented by a latent code map, denoted as $E_{\varphi}(M)$, where $E$ is the encoder, $\varphi$ represents its parameters, and $M$ is the texture image.
The latent code map stores feature information distributed in the local area of pixel coordinates within the image domain. By using query coordinates to query the latent code content, it can return the feature value at the query coordinate, such as the texture feature vector at a point on a triangular face of the current textured mesh.

\subsubsection{Implicit Interpolation}
Each triangle face on the textured mesh corresponds to an irregular triangular patch on the texture latent code map $E_{\varphi}(M)$, as shown in Figure \ref{fig:grid}.
To address this, we employ uniformly sampled receptive field convolution on each patch, producing a feature map of consistent scale for the irregular texture patches corresponding to each triangular face.
In other words, the convolution of the feature map patches is performed within the receptive field defined by the grid cells that uniformly sample the triangle.
To maintain a consistent scale for all patches without causing warping through adaptive density sampling, the grid cells used for feature extraction need to undergo coordinate transformation based on the shape of the triangles.
At the beginning, we expand the scale $(0,1)$ of UV coordinates to $(-1,1)$. Assuming the UV coordinates of the vertices of a triangular face are $T_0,T_1,T_2$, the horizontal and vertical bounds are $[x_{min},y_{min},x_{max},y_{max}]$.
To ensure that the receptive field covering the triangle during the convolution operation remains square, while maintaining consistent sampling density in both the horizontal and vertical directions, we position the triangle at the center and expand its horizontal or vertical range according to its aspect ratio, filling the extended area with neighboring feature map values. To avoid discontinuous feature estimation, we apply implicit interpolation using the values of the four nearest corners within the neighborhood for non-pixel feature points.
The implicit interpolation is implemented with voting and normalized confidence like bilinear method.
This implicit interpolation ensures a smooth transition even at coordinates that fall between pixel positions at the convolution grid points.
Then, the feature maps corresponding to each face are mapped onto the surfaces of the textured mesh, resulting in texture features with dimensions $F\times C$, where $C$ represents the number of channels.

\subsection{Geometric Extraction Module}
In a textured mesh, geometric features include spatial features that encode the geometric information of spatial points, structural features that describe the topology and overall shape of the mesh, and shape features that characterize the shape and irregularity of individual triangular faces.

\begin{figure}[!t]
  \centering
  \includegraphics[width=0.35\textwidth]{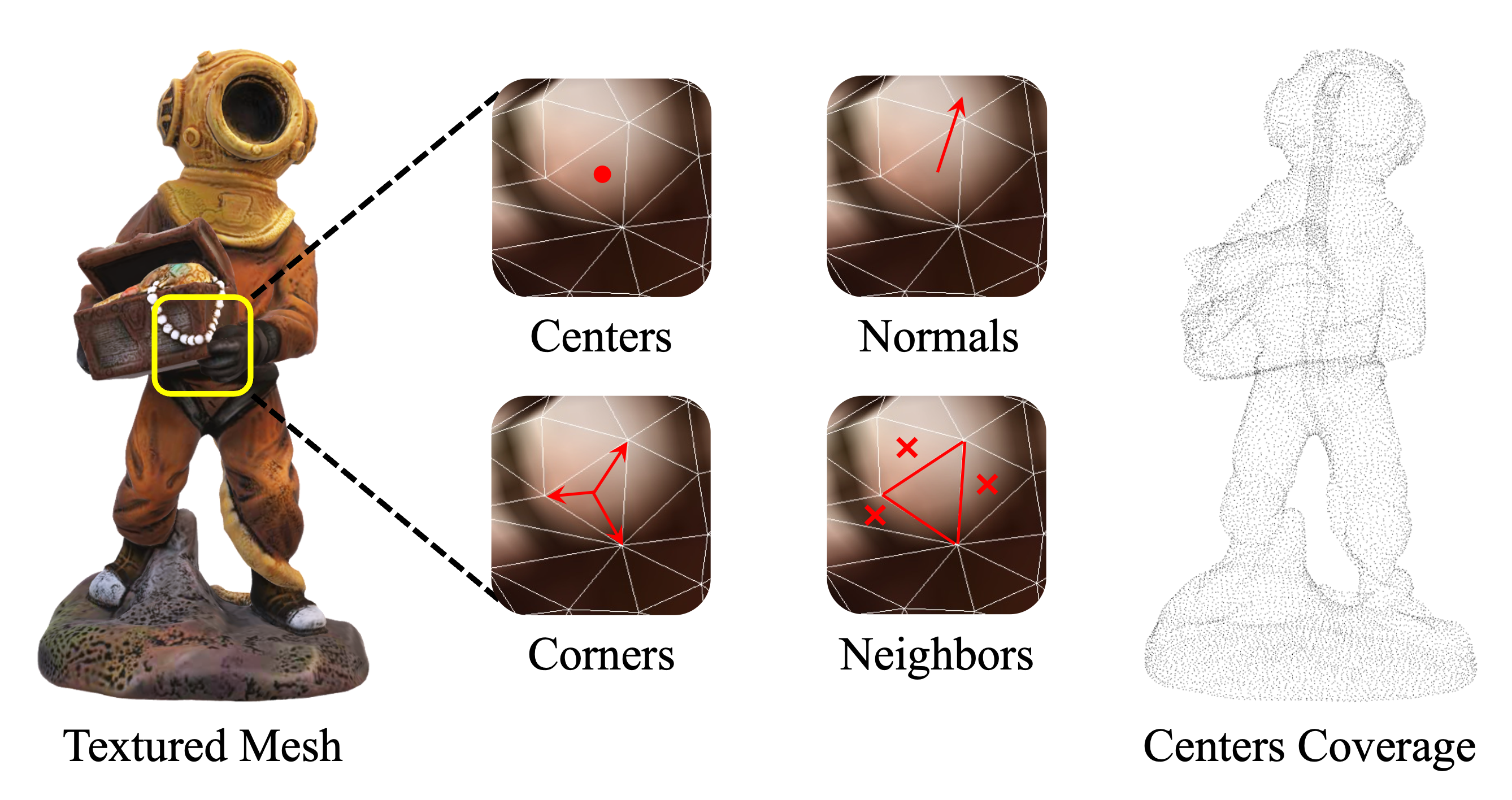}
  \setlength{\abovecaptionskip}{-0pt}
  \caption{Data types of geometric information. The face centers are sampled, and the centers coverage resembles that of point clouds.}
\label{fig:input}
\end{figure}
\subsubsection{Spatial Extractor}
The spatial extractor takes the center coordinates of the faces as input to extract each face's position in 3D space and the overall shape of the 3D object, illustrated in Figure \ref{fig:input}. The centers are typically provided as the $F\times 3$ matrix, where each center point has three coordinates $(x, y, z)$. This module uses a residual convolutional block to process the spatial coordinates, similar to point clouds, to output the spatial features of the 3D model.
Additionally, spatial coordinates of the entire 3D model are normalized and subjected to random rotations. These transformations improve the module's generalization capability while preserving the topological structure and connectivity.

\subsubsection{Shape Extractor}
Shape features refer to the irregularity of triangular faces on a textured mesh surface. In regions with fewer details, triangles may exhibit greater irregularity and sharper angles. Conversely, in areas with more detailed features, triangles tend to be more densely distributed and generally have more regular shapes. Thus, describing the shapes of triangles provides fundamental geometric information, and the complexity of the shapes is closely related to the distribution of saliency. The corners, represented by vectors from the face center to each vertex as shown in Figure \ref{fig:input}, are input as a $F \times 3 \times 3$ matrix. The angular relationships and lengths between these vectors reflect the shape of the triangles, which are then processed by the shape convolution block to extract shape features for each triangular face.

\subsubsection{Structural Extractor}
Structural features reflect the curvature of local regions on the surface. In textured mesh data, the face normals indicate the outward direction of the faces. The correlation between the normals of neighboring faces can measure the surface curvature. To enhance the receptive field of curvature features, we use 1-ring, 2-ring, and 3-ring neighboring faces as the neighboring range.
To capture the directional relationships of normal vectors, we use a set of learnable directional bases $V_K$ in the Euclidean coordinate system. Let $ N(f)$ denote the normal vector of the central face $f$, and $N_R(f)$ represent the normal vectors of its $R$-ring neighboring faces. Together, these vectors form the set $N(f, R) = \{N(f), N_R(f)\}$, which describes the local surface curvature. By calculating the maximum cosine similarity between the normal vector set $N(f, R)$ and the learnable directional bases $V_K$, the bases which best match the local structure can be identified, thereby more accurately representing the structural features of the 3D structure.

\subsection{Aggregation Module}
In the aggregation module, as shown in Figure \ref{fig:mod}, texture and geometric features are integrated to produce a comprehensive representation of the textured mesh. This module combines the detailed texture information extracted from the texture latent code with the geometric features derived from the mesh’s structure.
Using a cross-attention mechanism, the module can focus on relevant information from both texture and geometry. In the attention block, spatial and structural features are combined to calculate the attention weight matrix. 
In the projection block, spatial, structural, and texture features are folded and projected. Then, the fused features of each face are aggregated with those of its neighboring faces by rearranging the structural features based on local adjacency relationships, thereby enhancing the network's representation of the neighborhood.
Finally, the concatenated features are fed into a MLP to obtain the predicted saliency map.

\begin{table*}[!t]
\centering
\resizebox{0.9\textwidth}{!}{
\begin{tabular}{c|cccc|cccc|cccc}
\hline
                 & \multicolumn{4}{c|}{Geometry}                                          & \multicolumn{4}{c|}{Color}                                             & \multicolumn{4}{c}{Texture}                                            \\ \hline
Method Name      & CC $\uparrow$   & SIM $\uparrow$  & KLD $\downarrow$ & SE $\downarrow$ & CC $\uparrow$   & SIM $\uparrow$  & KLD $\downarrow$ & SE $\downarrow$ & CC $\uparrow$   & SIM $\uparrow$  & KLD $\downarrow$ & SE $\downarrow$ \\ \hline
PointNet         & 0.2409          & 0.6080          & 0.4963           & 0.0419          & 0.2390          & 0.6083          & 0.5046           & 0.0432          & 0.2281          & 0.6104          & 0.5097           & 0.0403          \\
PointNet2-SSG    & 0.3491          & 0.6459          & 0.4460           & 0.0362          & 0.3761          & 0.6481          & 0.4406           & 0.0375          & 0.3584          & 0.6414          & 0.4402           & 0.0372          \\
PointNet2-MSG    & 0.4749          & 0.6699          & 0.3903           & 0.0331          & 0.4280          & 0.6575          & 0.4120           & 0.0342          & 0.4389          & 0.6598          & 0.4088           & 0.0344          \\
PointTrans       & 0.5138          & 0.6821          & 0.3613           & 0.0311          & 0.4955          & 0.6756          & 0.3760           & 0.0308          & 0.5201          & 0.6817          & 0.3578           & 0.0297          \\
PointMixer       & 0.5261          & 0.6859          & 0.3531           & 0.0306          & 0.4929          & 0.6765          & 0.3755           & 0.0326          & 0.5246          & 0.6856          & 0.3576           & 0.0305          \\
StraTrans        & 0.5096          & 0.6798          & 0.3650           & 0.0304          & 0.4889          & 0.6703          & 0.3886           & 0.0315          & 0.5064          & 0.6756          & 0.3762           & 0.0314          \\ \hline
MeshNet          & 0.5512          & 0.6996          & 0.3382           & 0.0285          & 0.5535          & 0.6955          & 0.3411           & 0.0290          & 0.5605          & 0.7002          & 0.3371           & 0.0286          \\
MeshNet++        & 0.4167          & 0.6623          & 0.4104           & 0.0366          & 0.4389          & 0.6642          & 0.4064           & 0.0343          & 0.4183          & 0.6635          & 0.4012           & 0.0367          \\
DiffusionNet-xyz & 0.4351          & 0.6660          & 0.4219           & 0.0356          & 0.4198          & 0.6623          & 0.4226           & 0.0361          & 0.4296          & 0.6615          & 0.4222           & 0.0344          \\
DiffusionNet-hks & 0.3293          & 0.6065          & 0.4569           & 0.0397          & 0.3496          & 0.6333          & 0.4452           & 0.0388          & 0.3781          & 0.6393          & 0.4374           & 0.0365          \\
DSM\_CNN         & 0.2174          & 0.6186          & 0.4902           & 0.0408          & 0.2082          & 0.6108          & 0.4984           & 0.0422          & 0.2189          & 0.6022          & 0.5186           & 0.0465          \\
DSM\_FCN         & 0.2040          & 0.6160          & 0.5008           & 0.0414          & 0.2031          & 0.6096          & 0.5073           & 0.0450          & 0.2163          & 0.6038          & 0.5039           & 0.0445          \\
SAL3D            & 0.4446          & 0.6674          & 0.4008           & 0.0332          & 0.4206          & 0.6560          & 0.4297           & 0.0363          & 0.3588          & 0.6405          & 0.4562           & 0.0374          \\ \hline
Ours             & \textbf{0.5675} & \textbf{0.7043} & \textbf{0.3239}  & \textbf{0.0274} & \textbf{0.5571} & \textbf{0.6994} & \textbf{0.3408}  & \textbf{0.0283} & \textbf{0.5764} & \textbf{0.7061} & \textbf{0.3198}  & \textbf{0.0273} \\ \hline
\end{tabular}
}
\setlength{\abovecaptionskip}{2pt}
\caption{Quantitative performance at different stages: Geometry, Color, and Texture.}
\label{tab:res}
\end{table*}

\begin{figure*}[!htbp]
  \centering
  \includegraphics[width=0.90\textwidth]{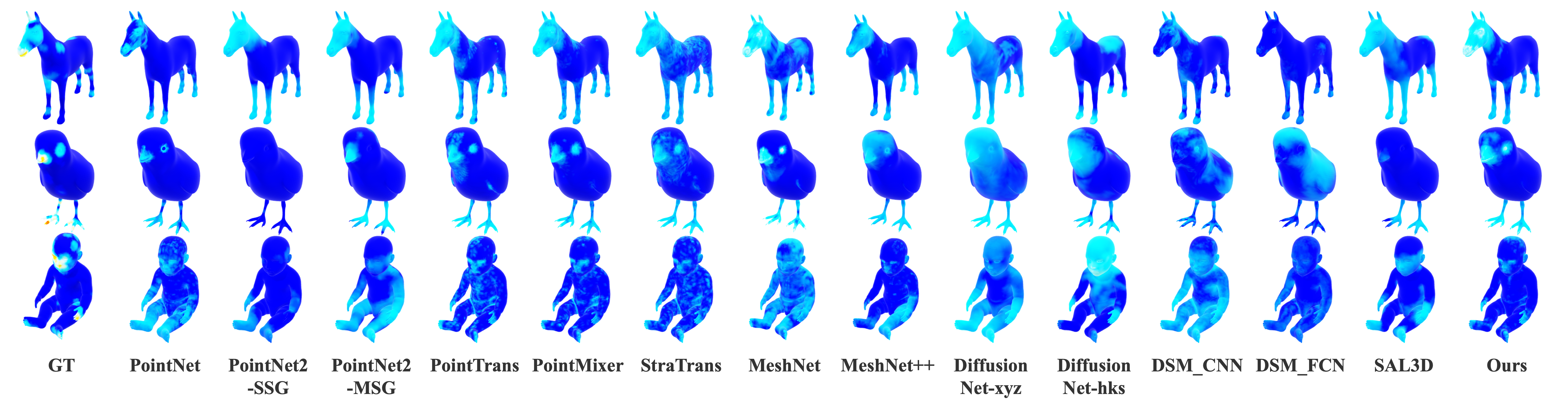}
  \setlength{\abovecaptionskip}{-0pt}
  \caption{Visualization results of compared methods at the Texture stage.}
\label{fig:tex}
\end{figure*}

\section{Experimental Results and Discussion}
Given that existing mesh saliency methods primarily focus on geometric features, we aim to extend our evaluation by integrating color and texture.
First, we compare various methods using only geometric structures. To ensure fairness, our model extracts geometric features and is compared against others under identical conditions. Next, we introduce vertex colors, enhancing our model with additional contextual information. For a balanced comparison, the same color extraction module is applied to all compared methods. Finally, we incorporate texture features, which improves model realism and accuracy. Again, to maintain fairness, the same texture alignment module is used across all methods for a consistent evaluation.
Since there are no other available datasets of sufficient scale for training, we decide to conduct the experiments on our built dataset.

\subsection{Experimental Setup and Implementation Details}
To more effectively quantify the prediction results, we employ Correlation Coefficient (CC), Similarity (SIM), Kullback Leibler Divergence (KLD), and Saliency Error (SE) calculated in the form of mean squared error as evaluation metrics.
We conduct our experiments using dual NVIDIA GeForce RTX 4090 GPUs and Intel i9 processor. For training all compared models, we use the L1-loss as the loss function. The AdamW optimizer is adopted for training all networks, with an initial learning rate set to 1e-3. This learning rate is reduced by a factor of 0.1 every 50 epochs. The training process for all methods is terminated after 250 epochs.
In order to evaluate the model's predictive performance, we conduct experiments on our self-constructed textured mesh saliency dataset. The dataset is divided into two parts: 80 samples for the training set, and the remaining 20 samples constitute the test set.

\subsection{Performance Evaluation}
\subsubsection{Baselines.}
The state-of-the-art 3D saliency deep models, DSM \cite{nousias2023deep} (configured with CNN-based and FCN-based approaches for 3D saliency patch descriptor) and SAL3D \cite{martin2024sal3d}, which are primarily used for saliency prediction tasks on non-textured meshes, will be used as baseline methods for performance comparison.
Furthermore, to address the lack of available comparison models, we adapt other 3D segmentation models by transforming the outputs into saliency prediction modules for a fair evaluation.
Specifically, since face center coordinates align with point cloud data, we select point-based models such as PointNet \cite{qi2017pointnet}, PointNet2 \cite{qi2017pointnet++}, PointTrans \cite{zhao2021point}, PoointMixer \cite{choe2022pointmixer} and StraTrans \cite{lai2022stratified} for this adaptation.
Subsequently, we employ mesh-based models that operate on faces as individual data units, adapting them for the saliency prediction task to ensure consistency.
The compared face-based models contain MeshNet \cite{feng2019meshnet}, MeshNet++ \cite{singh2021meshnet++}, DiffusionNet \cite{sharp2022diffusionnet} (the suffixes xyz and hks denote networks with positions and heat kernel signatures as input, respectively).

\subsubsection{Geometric Structure-Only Comparison.}
In this stage, we evaluate the performance of various methods based solely on geometric structures.
The quantitative results are shown in the Geometry column of Table \ref{tab:res}. Point-based methods generally perform moderately, while more advanced Transformer-based methods achieve better results. Additionally, methods like MeshNet that consider the mesh's topological structure show results closer to ours. In contrast, the DSM method, which is trained on patch-scale regions of mesh surface, performs less effectively. The latest SAL3D, being based on PointNet2, also shows a significant performance gap. Additionally, we present the visualization results in Appendix. As shown, our results display clear boundaries and effectively cover the main salient areas.

\subsubsection{Color-Enhanced Method Comparison.}
We introduce the same color extraction module to all methods, which allows for a consistent evaluation of how vertex color integration impacts performance.
The vertex colors here are mapped from the texture image to the vertex coordinates using UV mapping. As shown in the Color column of Table \ref{tab:res}, integrating vertex colors into our model results in a noticeable decrease in performance. Similarly, we present the visualization results in Appendix. Most of the other comparison methods also experience a decline in performance due to the inclusion of the vertex color information. This suggests that vertex colors have limited feature extraction capability for textured meshes, which exhibit complex visual effects due to different texture patterns.

\subsubsection{Texture-Integrated Method Comparison.}
We apply the same texture alignment module across all methods, which ensures a thorough comparison of the impact of texture mapping.
As shown in the Texture column of Table \ref{tab:res}, incorporating the texture alignment module significantly improves our model's performance. The saliency prediction visualization results are displayed in Figure \ref{fig:tex}, where our method also demonstrates a greater focus on key areas. Other methods that already perform well using only geometric features also show some improvement. However, point-based methods, which do not capture the mesh's topological structure, do not experience any performance enhancement. These results reflect the improved model performance achieved by integrating geometric structure with texture features.

\begin{figure}[!t]
  \centering
  \includegraphics[width=0.45\textwidth]{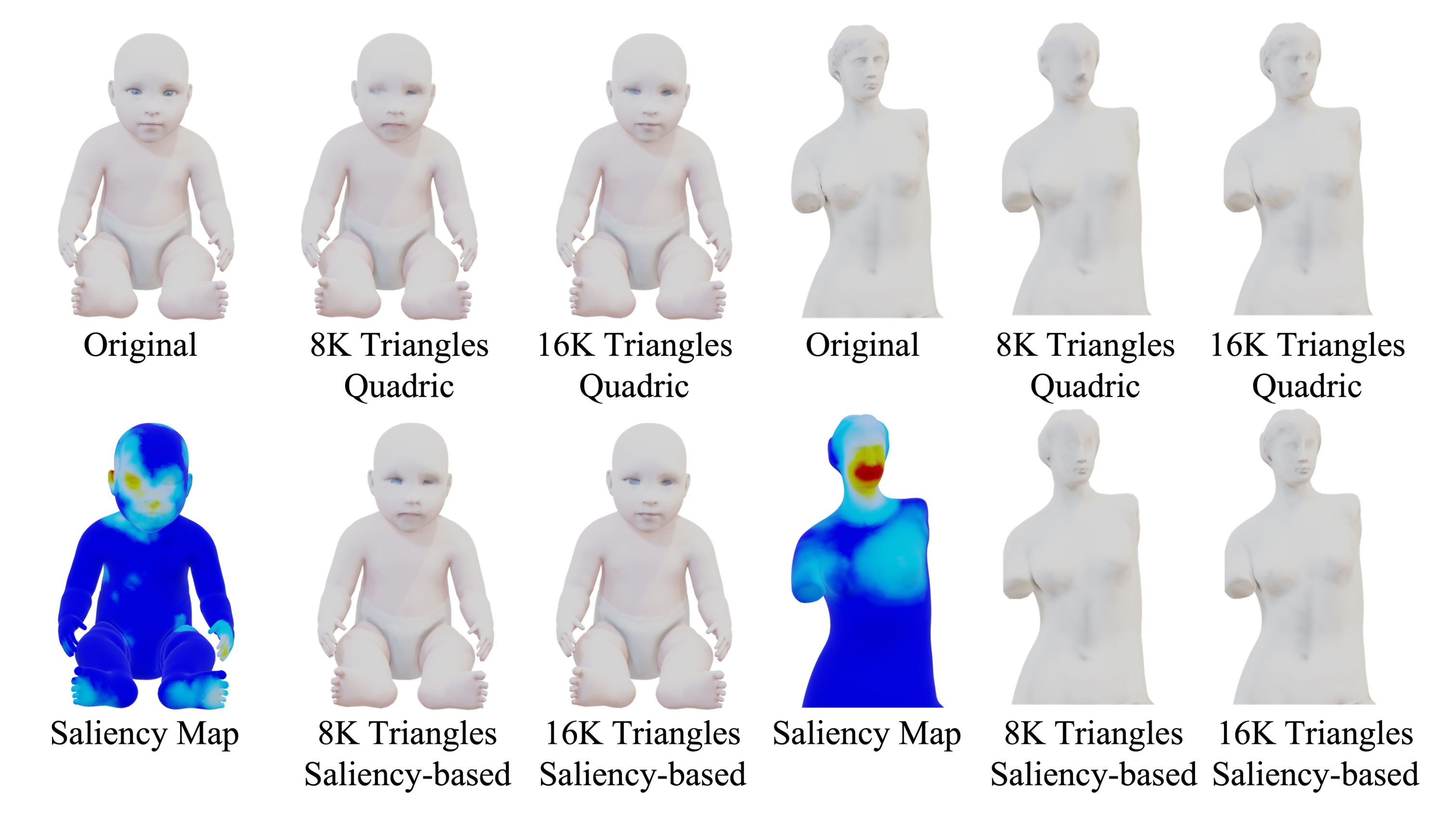}
  \setlength{\abovecaptionskip}{-10pt}
  \caption{Textured mesh simplification results with Quadric-based algorithm and saliency-based simplification.}
\label{fig:simplify}
\end{figure}

\subsection{Discussion on Saliency-Based Simplification}
When we observe a 3D object, our gaze tends to focus on unique details and salient features, such as distinctive textures or specific shapes. Areas on the object's surface where there are noticeable changes in color, lighting, or geometry also attract more attention. Regions with special semantic significance, like buttons or faces, which are often closely related to the object's function or category, are particularly likely to capture visual attention. These critical areas of a textured mesh cannot be fully captured by geometry and color textures alone, but saliency maps, modeled with human perception in mind, provide the model with enhanced sensitivity. 
In this block, we modify the classic Quadric-based Simplification algorithm \cite{garland1997surface} by incorporating saliency-based guidance and compare the results with the original algorithm. As demonstrated in Figure \ref{fig:simplify}, we present two examples with 8K and 16K triangle counts. The results clearly show that areas with high visual attention, such as the face, eyes, and torso, retain much more detail in the textured mesh.

\begin{table}[]
\centering
\resizebox{0.40\textwidth}{!}{
\begin{tabular}{c|cccc}
\hline
\# Subject      & CC $\uparrow$ & SIM $\uparrow$ & KLD $\downarrow$ & SE $\downarrow$ \\ \hline
w/o Texture     & 0.5675        & 0.7043         & 0.3239           & 0.0274          \\
w/o Spatial     & 0.5682        & 0.7016         & 0.3217           & 0.0275          \\
w/o Shape       & 0.4135        & 0.6639         & 0.4053           & 0.0361          \\
w/o Structral   & 0.5625        & 0.7013         & 0.3311           & 0.0278          \\
w/o Aggregation & 0.5585        & 0.7022         & 0.3286           & 0.0277          \\ \hline
\end{tabular}
}
\setlength{\abovecaptionskip}{2pt}
\caption{The performance of different combined components.}
\label{tab:abs1}
\end{table}

\begin{table}[]
\centering
\resizebox{0.45\textwidth}{!}{
\begin{tabular}{c|cccc}
\hline
\# Removed             & CC $\uparrow$ & SIM $\uparrow$ & KLD $\downarrow$ & SE $\downarrow$ \\ \hline
Latent Code            & 0.5530        & 0.6969         & 0.3534           & 0.0283          \\
Grid Expand            & 0.5637        & 0.7025         & 0.3287           & 0.0278          \\
Implicit Interpolation & 0.5708        & 0.7043         & 0.3243           & 0.0275          \\ \hline
\end{tabular}
}
\setlength{\abovecaptionskip}{2pt}
\caption{The performance of texture alignment module.}
\label{tab:abs2}
\end{table}

\subsection{Ablation Study}
In this section, we aim to validate the effectiveness of our proposed model through ablation experiments. We systematically remove or modify specific components and measure the resulting changes in prediction performance. This approach allows us to isolate the contributions of each component and gain insights into their relative importance.
We conduct experimental comparisons by separately removing the texture features as well as the spatial, structural, and shape features within the geometric features, with the results listed in Table \ref{tab:abs1}. It is evident that removing shape features has a catastrophic impact on saliency prediction, indicating a closer relationship between the distribution of triangular face shapes and saliency distribution. Additionally, to demonstrate the effectiveness of the aggregation module, we replace it with a direct concatenation of geometric and texture features.
In Table \ref{tab:abs2}, we validate the components of the texture alignment module through a series of tests. Initially, we remove the latent code and apply sampled convolution directly to the raw texture images. Next, we eliminate the expansion of the sampling grid. Finally, we replace implicit interpolation with nearest-neighbor sampling. The results demonstrate the effectiveness of this module.

\section{Conclusion}
In summary, our work addresses a critical gap in the field of 3D object saliency prediction by introducing a novel approach specifically designed for textured meshes.
Our innovative dataset, generated through novel eye-tracking experiments in a 6-DOF VR environment, offers comprehensive and diverse visual data that addresses the limitations of previous studies.
Our proposed model integrates a texture alignment module, a geometric extraction module, and an aggregation module, leading to more accurate predictions of saliency maps for textured mesh surfaces.
By aligning intricate texture features with the geometric structure of 3D models, our model significantly enhances the unified representation of color patterns and topological structures in texture meshes.

\section{Acknowledgments}
This work was supported in part by the National Natural Science Foundation of China under Grant 62271312, Grant 62132006, Grant 6213000376, Grant 62101325, Grant 62101326 and Grant 62377011; in part by the Shanghai Pujiang Program under Grant 22PJ1407400.

\bibliography{aaai25}

\end{document}